# Lattice QCD Thermodynamics on the Grid

Jakub T. Mościcki[1], Maciej Woś[1], Massimo Lamanna[1],
Philippe de Forcrand[2,3] and Owe Philipsen[4]

[1] *CERN, Information Technology Department, Grid Support, CH-1211 Geneva 23, Switzerland*
[2] *Institut für Theoretische Physik, ETH Zürich, CH-8093 Zürich, Switzerland*
[3] *CERN, Physics Department, TH Unit, CH-1211 Geneva 23, Switzerland*
[4] *Institut für Theoretische Physik, Westfälische Wilhelms-Universität Münster, Germany*

**Abstract**

We describe how we have used simultaneously $\mathcal{O}(10^3)$ nodes of the EGEE Grid, accumulating ca. 300 CPU-years in 2-3 months, to determine an important property of Quantum Chromodynamics. We explain how Grid resources were exploited efficiently and with ease, using user-level overlay based on Ganga and DIANE tools above standard Grid software stack. Application-specific scheduling and resource selection based on simple but powerful heuristics allowed to improve efficiency of the processing to obtain desired scientific results by a specified deadline. This is also a demonstration of combined use of supercomputers, to calculate the initial state of the QCD system, and Grids, to perform the subsequent massively distributed simulations. The QCD simulation was performed on a $16^3 \times 4$ lattice. Keeping the strange quark mass at its physical value, we reduced the masses of the up and down quarks until, under an increase of temperature, the system underwent a second-order phase transition to a quark-gluon plasma. Then we measured the response of this system to an increase in the quark density. We find that the transition is smoothened rather than sharpened. If confirmed on a finer lattice, this finding makes it unlikely for ongoing experimental searches to find a QCD critical point at small chemical potential.

# 1 Introduction

Quantum Chromodynamics (QCD) describes the strong interactions between quarks and gluons, which are normally confined inside protons, neutrons and other baryons. Because the interactions are strong, the analytic perturbative expansion, where one determines exactly the first few orders of a Taylor expansion in the coupling constant, converges poorly. Thus, one commonly resorts to large-scale Monte Carlo computer simulations, where the complete properties of QCD can be obtained numerically, up to controllable statistical and systematic errors.

In order to simulate QCD on a computer, one discretizes space and time into a 4-dimensional grid. The quark and gluon fields live respectively on the sites and bonds of this "lattice". The computer generates a sample of the most important configurations of quark and gluon fields, evolving them one Monte Carlo step at a time. Statistical errors come from the Monte Carlo sampling, and systematic errors come from the finite lattice spacing $a$ and finite size of the simulated 4-dimensional "box" $N_x a \times N_y a \times N_z a \times N_\tau a$. While the three space dimensions should be in principle infinitely large, the fourth dimension defines the temperature $T$ of the system: $T = 1/(N_\tau a)$.

The majority of lattice QCD simulations studies properties of the $T = 0$ theory. Thus, all four dimensions are large, and state-of-the-art projects with, say, $N_x = N_y = N_z \sim \mathcal{O}(32), N_\tau \sim \mathcal{O}(64)$, require distributing the quark and gluon fields over many CPUs, which must be efficiently inter-connected to maintain a reasonable efficiency. The accumulated statistics typically reach $\mathcal{O}(10^{3 \div 4})$ Monte Carlo "trajectories".

In contrast, we are interested in high-precision measurements of some properties of QCD at finite temperature. This means that the system we study, of size $16^3 \times 4$, fits into the memory of a single CPU, and that our large CPU requirements stem from the high statistics required, $\mathcal{O}(10^6)$ trajectories. In this case, a large pool of independent CPUs represents a cheap, efficient alternative to a high-performance cluster. This is why, in our case, using the EGEE Grid[1] was the logical choice.

The EGEE Grid is a globally distributed system for large-scale processing and data storage. At present it consists of around 300 sites in 50 countries and offers more than 80 thousand CPU cores and 20 PB of storage to 10 thousand users around the globe. EGEE is a multidisciplinary Grid, supporting users in both academia and business, in many areas of physics, biomedical applications, theoretical fundamental research and earth sciences. The largest user communities come from High-Energy Physics, and in particular from the experiments active at the CERN Large Hadron Collider (LHC).

The physics problem we address is the following. At high temperature or density, the confinement of quarks and gluons inside baryons disappears: baryons "melt" and quarks and gluons, now deconfined, form a plasma. When the net baryon density is zero, this change is a rapid but analytic ("smooth") crossover as the temperature is raised. On the contrary, at high baryon density it is believed to proceed through a true non-analytic, first order phase transition. This change of the nature of the transition as a function of baryon density or chemical potential is analogous to the one occurring in liquid-gas transitions as a function of pressure: at low pressure water boils, and in this first-order transition, it absorbs latent heat. With increasing pressure, the transition (boiling) temperature rises and the first order transition weakens, i.e. the latent heat decreases until it vanishes altogether at a critical point, where the transition is second order. Beyond this critical pressure, the transition to the gaseous phase proceeds continuously as a crossover (with no latent heat). Correspondingly in QCD, there may exist a particular intermediate baryon density where the latent heat of the QCD phase transition vanishes and the phase transition is second-order. The corresponding temperature and baryon density or chemical potential define the so-called QCD critical point, which is the object of both experimental and theoretical searches, the former by heavy-ion collision experiments at RHIC (Brookhaven) and soon at LHC (CERN), the latter by numerical lattice simulations.

---

[1] Enabling Grids for E-ScienE; http://www.eu-egee.org

In theoretical studies, one may also consider the $u, d, s$ quark masses as variable and investigate their influence on the order of the transition, as shown in Fig. 1 (left) as a function of $m_u = m_d$ and $m_s$ for zero baryon density. For small enough quark masses the phase transition is of first order and corresponds to a high-temperature restoration of the chiral symmetry, which is spontaneously broken at low temperature. This chiral phase transition weakens with increasing quark masses until it vanishes along a chiral critical line, which is known to belong to the $Z(2)$ universality class of the 3d Ising model [1, 2]. For still larger quark masses, the transition is an analytic crossover. At finite density, it is generally expected that the $Z(2)$ chiral critical line shifts continuously with $\mu$ until it passes through the physical point at $\mu_E$, corresponding to the critical point of the QCD phase diagram. This is depicted in Fig. 1 (middle), where the critical point of QCD with physical quark masses is part of the chiral critical surface. However, there is no a priori reason for this scenario. In principle it is also possible for the chiral critical surface to bend towards smaller quark masses as in Fig. 1 (right), in which case there would be no chiral critical point or phase transition at moderate densities. To decide between these two possibilities one must determine the chiral critical surface, which can be expressed near $\mu = 0$ as a Taylor series in $\mu/T$. For three equal-mass ("degenerate") quark flavours $m_u = m_d = m_s = m$, i.e. along the diagonal in Fig. 1, the critical quark mass $m_c$ as a function of $\mu$ can be expressed as

$$\frac{m_c(\mu)}{m_c(0)} = 1 + \sum_n c_n \left(\frac{\mu}{\pi T}\right)^{2n} , \qquad (1)$$

where only even powers appear due to an exact symmetry $\mu \to -\mu$ in QCD, and $c_1 > 0$ for Fig. 1 (middle), $c_1 < 0$ for Fig. 1 (right). For the case with a heavier strange quark one considers a fixed $m_s$ and Eq. (1) gives the dependence of $m_{u,d}^c(\mu)$ on $\mu$, i.e. a slice of the critical surface for that $m_s$. The goal of our study was to determine, via the measurement of $c_1$, which of the two scenarios Fig. 1 (middle) or (right) applies, for the case where the mass $m_s$ of the strange quark takes its physical value, and for a fixed, rather large ($N_\tau = 4$, i.e. $a \sim 0.3$ fm) lattice spacing.

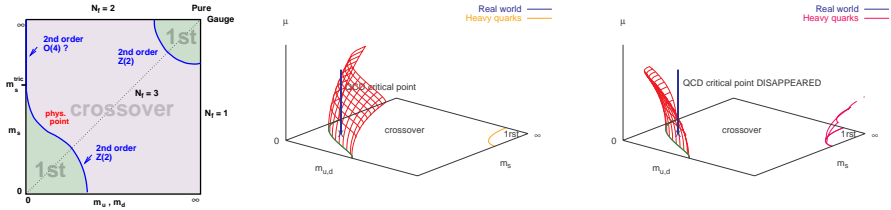

Figure 1: Left: Schematic phase transition behaviour of $N_f = 2+1$ QCD for different choices of quark masses $(m_{u,d}, m_s)$ at $\mu = 0$. The chiral critical line bounds the first order region in the lower left corner. Middle/Right: Critical surface swept by the chiral critical line as $\mu$ is turned on. Depending on the curvature, a QCD chiral critical point is present or absent. For heavy quarks the curvature has been determined [3] and the first order region shrinks with $\mu$.

What makes the project particularly interesting is that our earlier results obtained for the case of three degenerate quark flavours, $N_f = 3$, indicate that Fig. 1 (right) applies [4, 5], contrary to standard expectations. This surprising result needs to be confirmed for non-degenerate quark masses, which we do here, and on finer lattices. If it turns out to be a property of the continuum QCD theory, it will have a profound impact on the QCD phase diagram. In particular, it will make it unlikely to find a QCD critical point at small baryon density.

Our 4-dimensional "box" is defined on a $16^3 \times 4$ lattice[2]. The computer evolves the configurations of quark and gluon fields in a succession of Monte Carlo trajectories. The same system is studied at 18 different temperatures around that of the phase transition, and what is measured is the response to a small increase in the baryon density. The signal is tiny and easily drowned by the statistical fluctuations. The system is pre-thermalized on a supercomputer and then the necessary high statistics are obtained by running, at each temperature, many independent Monte Carlo threads. This strategy allows the simultaneous use of many Grid nodes in an embarrassingly parallel mode. In this way, the equivalent of 200-300 CPU-years was accumulated over a period of 2-3 months. Such a task requires approximately 1000 processors on the Grid running continuously over the entire period. To optimize the execution of the QCD simulations and to control the variance which occurs in very large Grids such as EGEE, we constructed a QCD production system using additional tools above the regular Grid middleware stack.

---

[2] We use the Wilson plaquette action for the gluons, and the staggered Dirac operator for the quarks, with masses $am_{u,d} = 0.005$ and $am_s = 0.25$.

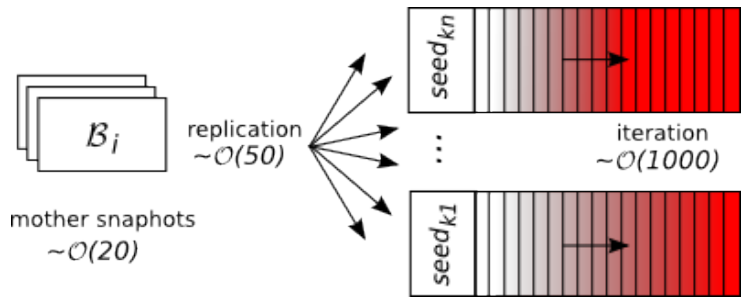

Figure 2: The high-level structure of LQCD simulation.

The details of our strategy are described in Sections 2-5, before we summarize the physics results in Section 6.

## 2 Computational model

The quark and gluon fields in the Lattice QCD simulation are mapped onto a discrete space-time lattice and evolved in Monte Carlo time. The system is studied at various temperatures which correspond to the values of the parameter $\beta$, the lattice gauge coupling constant. A complete lattice configuration is kept in a *snapshot* file and the initial configurations for each $\beta$-value are called *mother snapshots*. Each snapshot may be evolved in Monte Carlo time by a series of *iterations*. The signal to noise ratio is very small and requires a large number of iterations to become significant. However, if random number sequences are different then multiple parallel Monte Carlo threads may be used for the same $\beta$-value. The mother snapshots may be replicated and threads use different *random seeds*. The threads execute independently of one another. The snapshot's *maturity* is the number of iterations performed on that snapshot (see Fig. 2).

At first the replicas of the mother snapshots are identical. The subsequent iterations lead to randomization of the replicas. After a large number of iterations mature snapshots diverge enough to contribute statistically independent simulation results. Before the randomization point is reached the snapshots are *immature* and only provide statistically correlated contributions. The number of iterations needed to randomize the system was not a priori known and it was estimated to be between 300 and 500 iterations (corresponding to 20-30 CPU days) per snapshot. This corresponds to the amount of processor time "wasted" on randomization of the system. Therefore the problem does not obey typical scaling properties according to Gustafson's law but the speedup is limited by Amdahl's formula: the more parallel simulation threads, the more CPU time is wasted. Moreover if the number of available processors is variable and at some point smaller than the number of snapshots, a scheduling problem arises: how to choose a subset of snapshots in order to achieve a required number of "useful" iterations before the specified deadline? Processing more snapshots than the number of available processors would result in serialization of computations. Given the long randomization time and the large number of snapshots, a naive scheduling would lead to spending all CPU time in randomizing the replicas rather than doing "useful" work. Given the dynamic nature of the Grid environment, large fluctuations in the number of simultaneously running jobs were expected. Therefore the system was facing the following challenges:

- adapt the scheduling algorithm so that the number of "useful" iterations may be maximized;
- manage the utilization of resources available in the Grid on par with the number of parallel simulation threads;
- run autonomously over long periods of time.

# 3   Production

## 3.1   Implementation

The QCD production system is based on a scheduler capable of implementing application-aware policies and acting upon a dynamic pool of EGEE worker agents. We exploit the pilot-jobs pattern using DIANE [3] — a user-level master-worker framework[6] and Ganga[4] — a job management interface [7]. The system consists of three main components as shown in Fig. 3: the master, the worker agents and the submission tool.

The master is responsible for task scheduling and controls the order in which the snapshots are scheduled for processing to the individual workers. The snapshot files are stored on the local file-system of the master and are exchanged with the worker nodes using the DIANE file transfer service. Small application plugins written in the Python programming language were used to customize the framework for the needs of the Lattice QCD production.

Each worker performs a given number of iterations and uploads the resulting snapshot file back to the master. The snapshot is then ready to be evolved further by a free worker agent. In order to avoid unnecessary network traffic, once a particular worker agent downloads a snapshot, it keeps processing it as long as possible. Therefore the snapshot does not have to be downloaded multiple times and the worker continues the simulation using the snapshot already cached at the worker node. The worker agent runs as a Grid job and it has limited life time. The typical limiting factor is the time limit on the batch system at the Grid sites.

Workers are submitted as Grid jobs using the Ganga interface. This may be done manually by the user or in automatic mode, where the submission is controlled by the Agent Factory which provides a continuous stream of workers (see 5).

---

[3] DIstributed ANalysis Environment, http://cern.ch/diane
[4] http://cern.ch/ganga

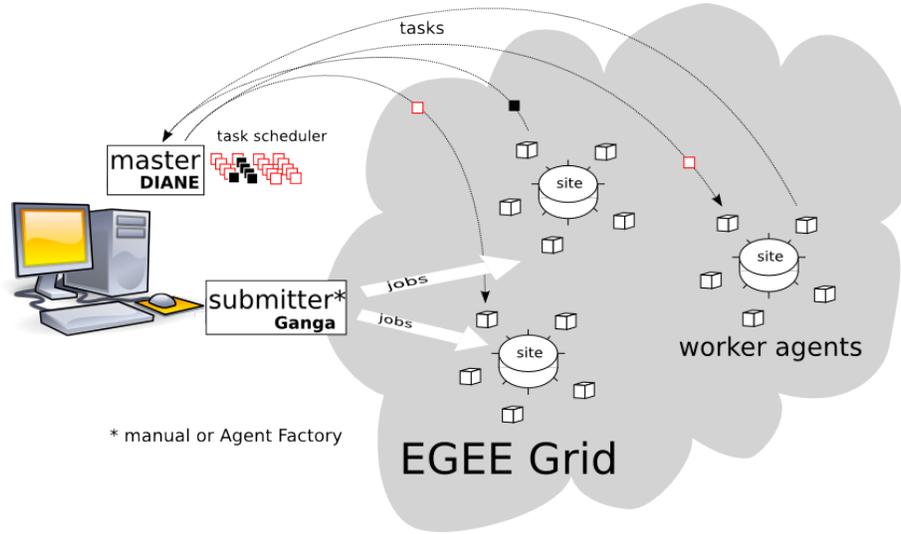

Figure 3: The LQCD production system architecture.

| Peak Performance | 1.2 TFlops |
| --- | --- |
| Processors | 80 CPUs |
| Number of Nodes | 10 |
| Memory/node | 128 GB |
| Disk | 160 TB shared disk |
| Node-node interconnect | IXS 8 GB/s per node |

Table 1: NEC-SX8 supercomputer characteristics.

## 3.2 Operation

The pre-thermalization of the Lattice QCD system was performed on a NEC-SX8 vector machine (Table 1) at HLRS in Stuttgart [5]. About 10 CPU minutes were required per Monte Carlo trajectory, and about 500 trajectories per $\beta$-value were produced. The fundamental reason for using vector machines in the pre-thermalization phase is the considerably higher throughput than the average node on the Grid. As finer lattice spacings are involved and the lattices get larger, exploiting fine-grained parallelism may also be beneficial. In this case a parallel architecture with low-latency interconnect is required.

The mother snapshots obtained on a NEC-SX8 vector machine were then used for the subsequent production on the EGEE Grid which took place from from April to October 2008 (and then followed by additional runs).

The production on EGEE Grid was split in several runs. The production workflow (Fig. 4) involved an active participation of the end-users: the inter-mediate simulation results were analyzed on-the-fly by the end-user scientists. This lead to several modifications and fine-tuning of the production including the simulation code, the number of $\beta$-values, the number of snapshots and the scheduling algorithms. The production was also interrupted due to technical reasons such as service upgrades or hardware downtime. The production phases are summarized in Table 2.

---

[5]Nec-SX8 technical description, http://www.hlrs.de/systems/platforms/nec-sx8

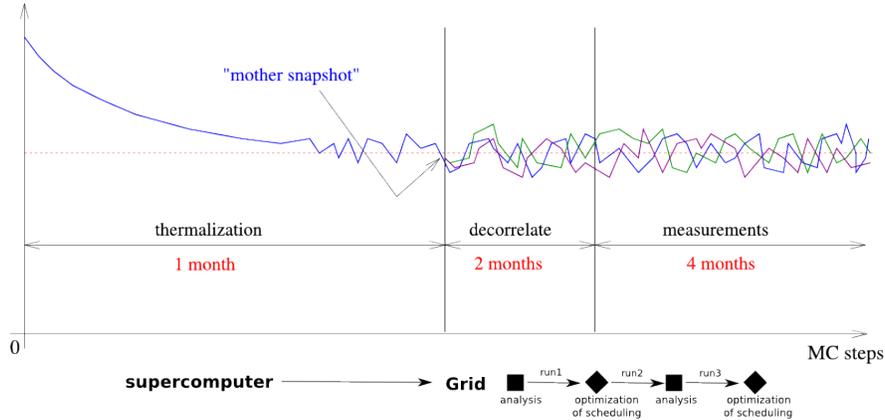

Figure 4: Monte Carlo time history of a typical measured QCD observable, illustrating the LQCD production workflow.

The goal of runs 1 and 2 in the first, most critical phase of the production, was to achieve 700,000 iterations, including the snapshot randomization, within approximately 10 weeks, in order to obtain publication-quality results. The average execution time per iteration was estimated at 1.5-2.5 CPU-h on a typical PC. The size of the snapshot file (input and output of each iteration) was 10 MB. In run 1 the system was analyzed for a quark mass of 0.0065 with 16 $\beta$-parameters (i.e. temperatures) uniformly distributed in a value range $(5.1805, 5.1880)$. In run 2 the estimation of the quark mass was refined at 0.0050. Additional $\beta$-parameters were defined in the middle of the range and placed in between the existing values to provide more simulation points in the vicinity of the phase-transition point to the quark-gluon plasma, further referred to as *sensitive region*. The reduction of the quark mass lead to longer execution time for the Monte Carlo step. This was compensated by reducing the frequency of measurements, to obtain a small overall reduction in CPU time per iteration. Run 1 and 2 were performed in parallel. The simulation parameters of run 2 were better tuned, therefore run 2 has eventually become the reference for the publication of physics results [8]. Production runs 3 and 4 were performed in a second phase and provided more precise data for subsequent studies. The $\beta$-value range was reduced as well as the total number of snapshots.

| run | $N_\beta$ | $N_{snapshot}$ | duration [weeks] | iterations [$\times 10^3$] | $N_{CPU}$ | $T_{CPU}$ [years] | data transfer [TB] |
|---|---|---|---|---|---|---|---|
| 1 | 16 | 400 | 11 | 300 | 4142 | 52 | 1.4 |
| 2 | 24 | 1450 | 9 | 700 | 21432 | 121 | 3.4 |
| 3 | 18 | 1063 | 3 | 267 | 12197 | 47 | 1.3 |
| 4 | 18 | 1063 | 8 | 266 | 12105 | 59 | 1.3 |
| total | | | 31 | 1533 | 49876 | 279 | 7.5 |

Table 2: Summary of LQCD production runs

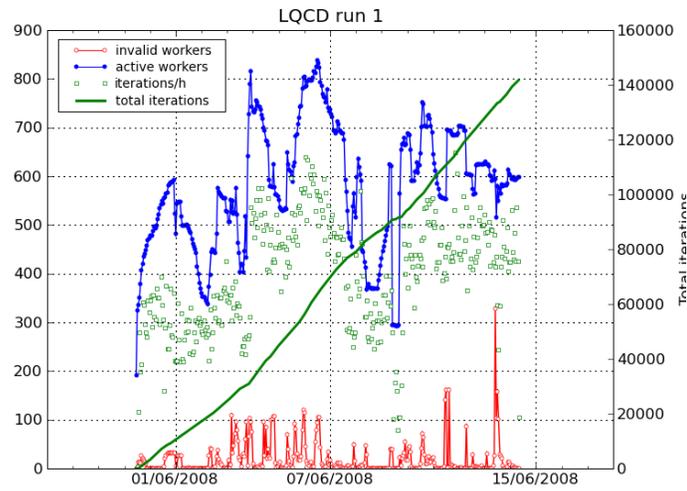

Figure 5: History plot showing the evolution of processing and of the worker pool size in run 1 (selected period). Manual submission of worker agents. Detailed explanation - see text.

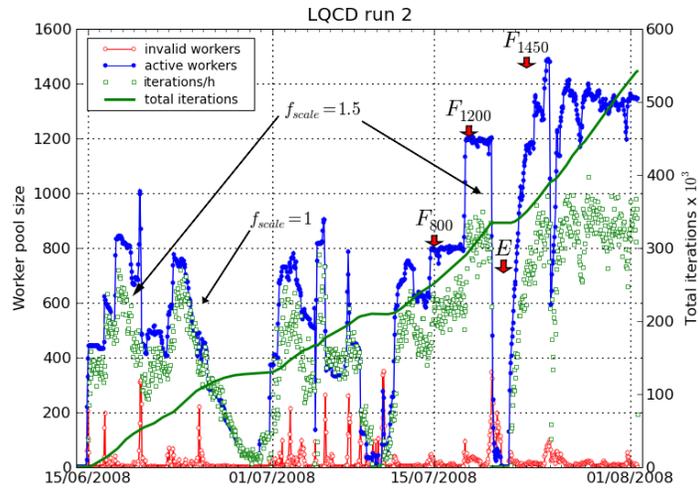

Figure 6: History plot of run 2 (selected period). Manual submission until 15/07. Meaning of symbols: $F_N$ – indicates the moment when the factory was enabled to keep N worker agents in the pool; $E$ – workers dropped due to expired user Grid credentials; $f_{scale}$ – rescaling factor for the number of iterations per hour (see text).

## 3.3 Analysis of system performance

The system analysis is based on the monitoring data collected by the DIANE master. For each run a journal file is generated which contains a complete record of events that occur between the master and the workers, and which is used to extract system parameters such as the number of active workers, the number of added workers, task duration etc. All the quantities have been sampled in one hour intervals.

The evolution of run 1 is presented in Fig. 5, run 2 in Fig. 6, run 3 in Fig. 7 and run 4 in Fig. 8. Due to missing data only selected periods of each run are shown. The left vertical axis shows the size of the worker pool i.e. the number of worker agents and the number of produced iterations in each time interval. The right vertical axis provides the scale for the total number of iterations. Some exceptional events occuring during the runs, such as expiry of Grid user credentials or server problems, add to natural fluctuations of the worker pool. The most important events are marked with arrows and described in the captions.

The lifetime of worker agents is limited by the batch queue limits and therefore the pool of productive workers is constantly changing. The workers which run the simulation during the time interval and successfully upload the result later are considered *active*. Workers which run the simulation but which do not upload the result are not considered active in a given time interval. This is the case of workers which were interrupted by the batch system due to the batch system time limits. In practice every worker becomes eventually inactive for a certain time before termination: a worker gets the workload from the master and runs the simulation which is interrupted by the batch system when the time limit is exceeded. This effect is called premature worker cancellation. The workers which never became active, i.e. did not manage to upload any results at all, are considered *invalid*. This may happen if, for instance, the simulation program cannot be started due to wrong processor architecture.

The number of performed iterations in a given time interval is proportional to the number of completed snapshots by the active workers in the pool. Each snapshot is uploaded after 3 completed iterations. The ratio between the number of active workers and the number of produced snaphots per hour is indicated by $f_{scale}$ on Fig. 6: $f_{scale} \simeq 1.5$ is a typical value for most of the runs while $f_{scale} \simeq 1.0$ corresponds to a larger number of faster workers being available in the Grid in certain periods.

## 4 Scheduling

In this Section we describe the application-specific scheduling which was applied to the LQCD production system. We use the inside knowledge of the $\beta$-parameter space to improve the total simulation throughput by ranking and prioritization of tasks to maximize the scientific content of the simulation output.

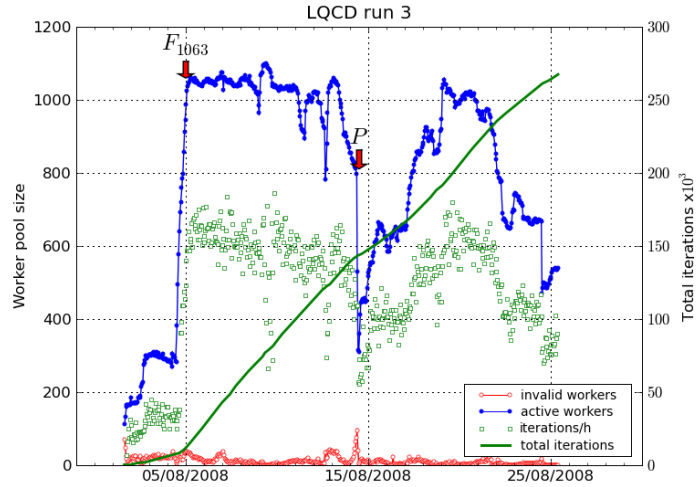

Figure 7: History plot of run 3 (selected period). Meaning of symbols: $F_{1063}$ – factory enabled for 1063 workers in the entire period; $P$ – power failure of the master server.

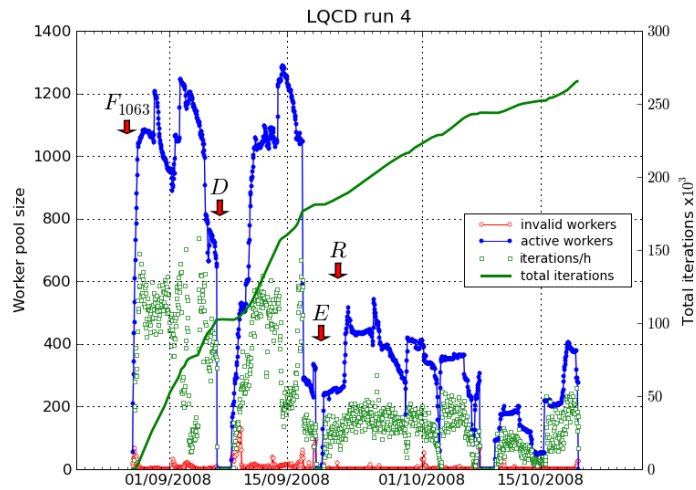

Figure 8: History plot of run 4 (selected period). Meaning of symbols: $F_{1063}$ – factory enabled for 1063 workers in the entire period; $D$ – file-server running out of file descriptors, system halted; $E$ – workers dropped due to expired user Grid credentials; $R$ – beginning of a period of low resource availability in the Grid, system working in low regime.

The computational complexity of the simulation (the average execution time) decreases when the temperature of the quark-gluon plasma increases (higher $\beta$ values). This effect is related to the physical behaviour of the system across the transition temperature. The theoretical curve of the distribution of the execution time as a the function of $\beta$ should be S-shaped and monotonically decreasing, with the inflection point at the phase-transition temperature.

Fig. 9 shows the execution time of iterations for each run. The vertical bars show the parameter range between the 25% and the 75% percentile. The points show the value of the median (50% percentile). The absolute values for each run differ because the internal parameters of the simulated system were modified between the runs. The rather large vertical bars account for relatively broad distributions. These distributions are dominated by the variability of computing resources which do not allow to further constrain the transition temperature using the execution time information. However two categories may be distinguished: $\beta < 5.1818$, where the QCD system is at low temperature, characterized by many small eigenvalues of the Dirac matrix and hence a larger computing time in the associated linear solver, and $\beta > 5.1820$ at high temperature, where the plasma is formed and the small Dirac eigenvalues disappear. We also expect a secondary peak of the other kind in each distribution because the separation into two categories is valid only for an infinitely large system. This is confirmed by the task execution histogram presented in Fig. 10 which shows a secondary peak at small CPU-time in the low-$\beta$ category.

At each temperature, the observed distribution of execution times results from the convolution of the distribution of the amount of computation required with another distribution reflecting the variability of Grid computing resources. We tried to disentangle these two distributions by first considering the highest temperature, where the amount of computation required fluctuates the least. A reasonable description of the distribution of execution times $t$ is obtained by the empirical function $(t/t_0-1)^{3/2}\exp(-3t/t_0)$ where $t_0$ represents the intrinsic, minimum execution time, which is smeared into a broad distribution including an exponential tail by Grid variability. Deviations from such a form at lower temperatures would then allow us to reconstruct the intrinsic distribution of $t_0$. Unfortunately, the fit remained rather good at all temperatures $\beta$, with a single parameter $t_0(\beta)$, showing that the observed distribution was mostly caused by Grid variability. The fitted intrinsic time $t_0(\beta)$ increases monotonically as the temperature is decreased, with a steepest variation near the critical temperature, as expected on physical grounds.

## 4.1 Maturity-based scheduling

In run 1 the snapshots were dynamically prioritized based on their maturity: the snapshots with the least number of Monte Carlo iterations were scheduled before the more mature ones. Let $S^s_\beta(k)$ denote a snapshot after $k$ iterations at a given temperature $\beta$ and for initial random seed $s$. Let $S^{s_1}_{\beta_1}(k_1) < S^{s_2}_{\beta_2}(k_2)$ denote that $S^{s_1}_{\beta_1}(k_1)$ should be scheduled before $S^{s_2}_{\beta_2}(k_2)$. The maturity-based

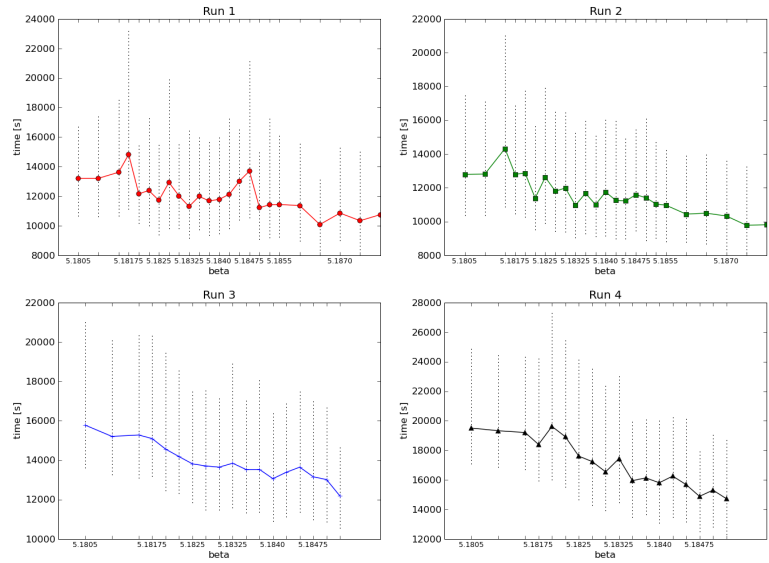

Figure 9: Median of execution time of 3 iterations for each $\beta$-value. The vertical bars show the parameter range between the 25% and the 75% percentile. The points show the value of the median (50% percentile).

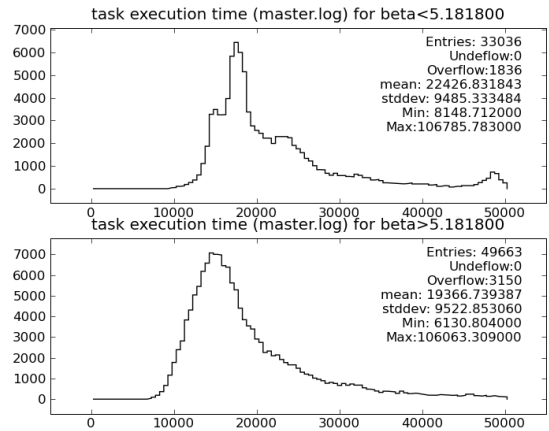

Figure 10: Distribution of task execution times in run 4 for two states of the QCD system: low temperature (top) and high temperature (bottom).

scheduling policy may be defined as:

$$S^{s_1}_{\beta_1}(k_1) < S^{s_2}_{\beta_2}(k_2) \iff k_1 < k_2 \qquad (2)$$

In first approximation, the objective is to evolve all snapshots keeping the spread in the iteration number as small as possible.

## 4.2  Scheduling in the sensitive region

After the initial analysis of the results, it was decided by the user to change the range of $\beta$-parameters and the scheduling policy for run 2 to improve the convergence speed. A finer-grained sensitive $\beta$-region $R = [5.1815, 5.18525]$ around the expected plasma transition temperature was defined. Within the sensitive region $R$ the scheduling policy was to select snapshots with a smaller $\beta$-value first

$$S^{s_1}_{\beta_1}(k_1) < S^{s_2}_{\beta_2}(k_2) \iff \beta_1 < \beta_2 \qquad (3)$$

Outside of the sensitive region the maturity-based prioritization was kept. Snapshots from the sensitive region were always selected before any snapshots from the outside of the region. Thus the scheduling policy in the entire range was defined as

$$S^{s_1}_{\beta_1}(k_1) < S^{s_2}_{\beta_2}(k_2) \iff \begin{cases} \beta_1 \in R\,,\ \beta_2 \notin R & or \\ \beta_1 < \beta_2 \text{ and } \beta_1, \beta_2 \in R & or \\ k_1 < k_2 \text{ and } \beta_1, \beta_2 \notin R & \end{cases} \qquad (4)$$

This gives absolute priority to the sensitive region, and within that region to smaller $\beta$ values.

In runs 3 and 4 the sensitive region was expanded to include $\beta$-values below the expected phase transition point, $R = [5.1805, 5.18525]$. At the same time the $\beta$-values above the sensitive region were removed from the simulation.

## 4.3  Analysis of the scheduling results

Fig. 11 shows the maturity of the snapshots at the end of each of the runs, grouped by the values of $\beta$. The final distribution of maturity depends on the computational requirements for each $\beta$-value, the scheduling policy and runtime factors.

Results obtained for run 1 and run 2 show that the maturity-based scheduling is implemented efficiently. However the final maturity distribution in the sensitive region is influenced by number of available processors in the system.

Scheduling in the sensitive region is based on the ordering of the snapshots (with respect to the $\beta$ parameter). Therefore, in first approximation the number of completed iterations per snapshot should be larger for smaller $\beta$-values. This applies to a system working at low regime, i.e. when the number of available processors is smaller than the number of snapshots. The effect is observed for run 4 in Fig. 8: more than half of the time the system is working at 500 workers or less, what corresponds to 50% of the required processing capacity.

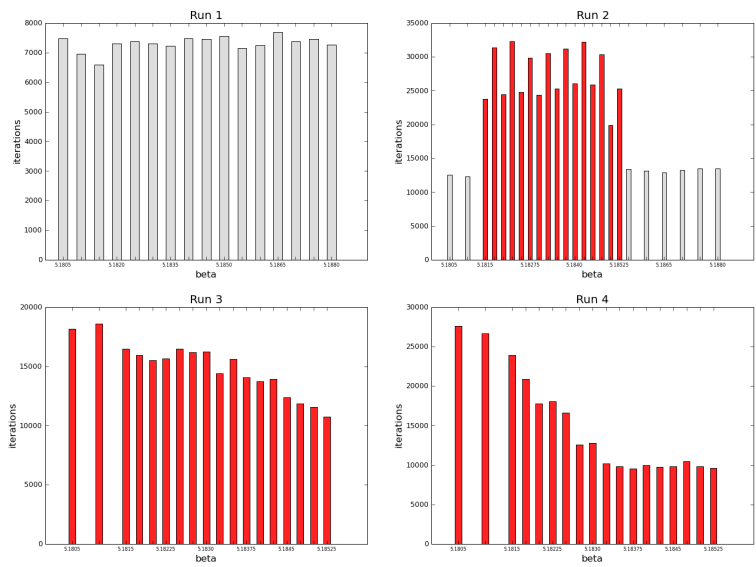

Figure 11: Total number of completed iterations in the different beta ranges for all runs. The sensitive region is indicated in darker colour (red). The sawtooth pattern visible in run 2 is due to interleaved $\beta$-values added after the run started, thus completing less iterations.

At full capacity, the number of available processors is equal to or higher than the number of snapshots. In this case the actual ordering priority does not matter because each snapshot is processed at any time. Considering that the worker pool is constantly changing (workers join and leave quite often) the ordering of snapshots does not influence the effective allocation of snapshots to workers. The final maturity of snapshots at the end of a run depends on the amount of processing in the function of $\beta$ and the distribution of processing power of the workers.

## 5    Resource provisioning by the Agent Factory

The aim of resource provisioning is to maintain the number of active workers as high as possible but not greater than the total number of snapshots. As an example, in case of a sudden drop in the number of workers one would like to react by submitting new workers to replenish the worker pool. On the other hand the submission of new worker agents should be kept under control and on par with the number of available resources in the Grid at a given moment. In particular, if not enough free resources are available, then an excessive, continuous submission would have little effect on the speedup of the system but could lead to overloading of Grid services.

In the initial production run the workers were submitted by hand by the users, which required frequent manual operation and was time consuming and inefficient. The Agent Factory component was designed to automate the submission process and to optimize the resource selection, based on system feedback using recent performance data. This is accomplished through a resource selection algorithm in which the Grid sites are ranked based on their reliability and performance over time. The algorithm is designed to cope with the observed dynamics of the Grid where the number of available computing resources is variable in time. Typical cases include sites entering downtime periods and stoping accepting jobs, or sites going back into production after configuration fixes, hardware upgrades etc. The selection algorithm gives more weight to more recent performance data and eventually "forgets" old data. It allows to increase the submission success rate and to maintain the number of worker agents on a predefined level.

### 5.1    Selection algorithm

In EGEE Grid the sites or resources within a single cluster or batch farm are represented by *Computing Elements* (CEs). CE is the smallest management unit for the resource selection algorithm. The Agent Factory submits jobs to the CEs via the *Workload Management System* (WMS) which is used as a gateway to the Grid.

The core of Agent Factory is a non-deterministic selection procedure based on a fitness algorithm commonly found in genetic algorithms/evolutionary strategies [9]. When a new worker agent is submitted to the Grid, a CE is chosen randomly with probability proportional to its fitness. CEs are selected by the Agent Factory and the WMS simply forwards the job submission requests.

For a Computing Element with $n$ total jobs, $r$ jobs currently running and $c$ jobs completed without errors, the $fitness$(CE) is defined as

$$fitness(CE) = \frac{r+c}{n} \quad (5)$$

The fitness value lies in the [0..1] interval. The $fitness = 1$ represents a reliable CE with all workers either running or finished cleanly. If all workers are queuing in a CE or if a CE is unable to correctly execute any jobs for our application then $fitness = 0$.

For each CE the probability to be selected by the Agent Factory is

$$P(\text{CE}) = \frac{fitness(\text{CE})}{1 + \sum fitness(\text{CE}_i)} \quad (6)$$

The denominator is the total fitness of the population of known CEs, where 1+ addend represents a *generic CE slot* which corresponds to a random CE selected by the WMS. The generic slot is used for discovery and adaptive ranking of CEs. At bootstrap the list of known CEs is empty and all jobs are submitted via the generic slot. As the list of known CEs grows the Agent Factory keeps on using the generic slot to submit a small fraction of jobs to random sites to detect the availability of new resources or an improvement in the performance of CEs with low fitness.

The added value of the Agent Factory is that it ranks available resources as a function of the current performance for the *specific application* we are running. The Agent Factory is an efficient way to automatize resource provisioning without overloading the system with unnecessary submissions, hence to maximize the overall duty cycle of our application.

## 5.2 Analysis

In run 1 (Fig. 5) and in the first part of run 2 until 15/07 (Fig. 6), the workers were added to the pool with manual job submission by users without adhering to any particular submission schedule. In the remainder of run 2 and in runs 3,4 (Figs. 7, 8), the Agent Factory was enabled to maintain $N$ active workers in a pool as indicated by $F_N$ events. When Agent Factory is enabled, the number of invalid workers is less scattered and under better control. The resource selection algorithm of the Worker Factory reduces the number of invalid worker agents and thus reduces the number of failing jobs flowing in the Grid system which have negative impact on scheduling due to premature worker cancellation as described in Section 3.3. However a small background of invalid workers exists and it is a feature of the selection agorithm were a fraction of jobs are submitted to random CEs via the generic slot.

In the case of exceptional events, such as expiry of Grid credentials ($E$), the number of invalid workers rapidly increases as the number of active workers falls sharply. The number of compatible resources suddenly drops to zero as all new submissions fail and all running workers are interrupted. Such events have a similar impact on the system, independently of whether the workers are submitted manually or via the Agent Factory.

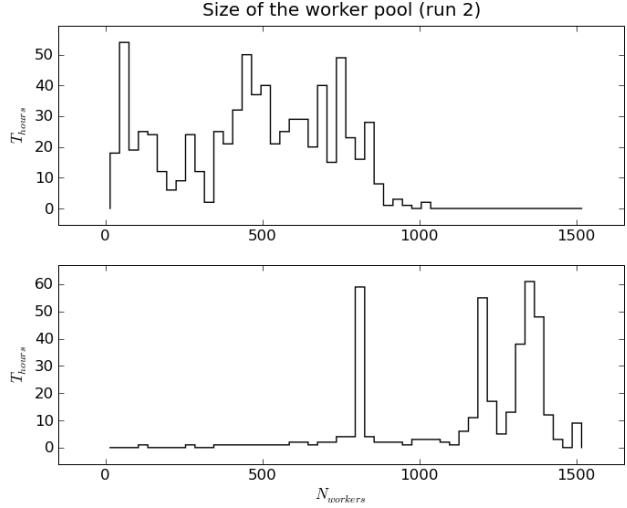

Figure 12: Distribution of the number of active workers in run 2 for submission without (top) and with (bottom) Agent Factory.

The distribution of the worker pool size in run 2 is shown in Fig. 12. In the manual submission mode, the distribution shows a large scatter below the optimal threshold for $N_{workers} = N_{snapshots} = 1450$, which indicates under-provision of the worker agents to the system. In case of Agent Factory, the three clear peaks of the distribution correspond to three stages of the run, as indicated by the events $F_{800}$, $F_{1200}$ and $F_{1450}$. When Agent Factory is enabled and resources are available, then the number of active workers quickly converges to the requested level and is maintained for an extended period of time.

The Agent Factory may not maintain the required level of the workers in the pool if there are not enough resources in the Grid. In run 4, the production enters a low regime phase $R$, where the amount of available resources is clearly below the optimal target of $N_{workers} = N_{snapshots} = 1063$. Under such conditions the resource selection based on best fitness allows to reduce the number of invalid workers as compared to the manual submission.

Occasionally the Agent Factory leads to over-submission of worker agents, e.g. $F_{1063}$ in run 4. The default policy of the Agent Factory is to fill up available computing slots within a Computing Element until the worker agents start queuing. If many computing slots become available at the same time in a larger number of Computing Elements, then a large number of queuing jobs suddenly start running and the worker pool could grow beyond the requested

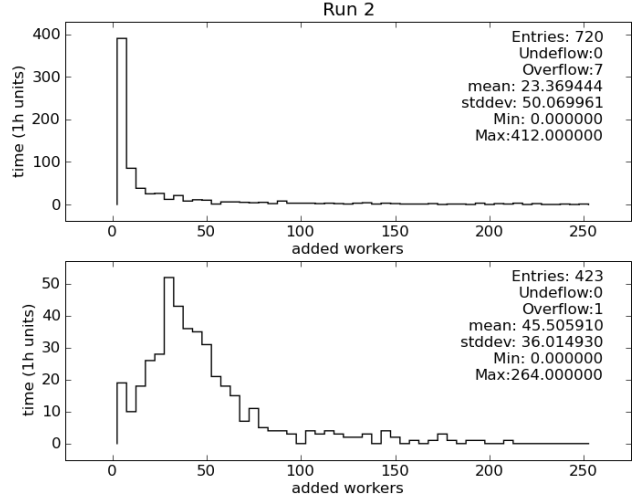

Figure 13: Distribution of the number of added workers per time unit in run 2 for submission without (top) and with (bottom) Agent Factory.

size.

The Agent Factory submits workers more efficiently such that a larger number of active workers is added to the system in a time unit. In run 2, the Agent Factory yields a stream of 43 active worker agents per hour as compared to 23 active worker agents per hour in manual submission mode. The distribution of the number of added workers in a unit of time, shown in Fig. 13, displays a clear difference in the submission patterns.

Finally, the Agent Factory allows the system to work autonomously and we observed a drastic cut in the time needed for human operation. Only seldom incidents, such as power cuts, needed manual interventions.

## 6 Physics results

With the computational model described in the previous Sections, we obtained a total of 1.5 million Monte Carlo trajectories, distributed over 18 $\beta$-values. As described in Section 1, we now wish to calculate the curvature of the chiral critical surface at the physical strange quark mass from these. An observable allowing to do this is the Binder cumulant

$$B_4 = \frac{\langle \delta X^4 \rangle}{\langle \delta X^2 \rangle^2} \;, \qquad (7)$$

where $\delta X = \bar{\psi}\psi - \langle \bar{\psi}\psi \rangle$ denotes the fluctuation in the order parameter for the chiral phase transition of interest, the chiral condensate $\bar{\psi}\psi$. In the infinite volume limit the Binder cumulant, evaluated on the phase boundary, i.e. at the critical coupling $\beta_c(m_{u,d})$, specifies the order of the phase transition by assuming values of 1 and 3 in a first order and crossover regime, respectively, and 1.604 for a second order transition in the $3d$ Ising universality class. On a finite lattice this step function is approached by a smooth analytic function changing more rapidly with increasing volume. The curvature of the critical surface, which is our quantity of interest, is directly related to the change of $B_4$ with chemical potential and quark mass. In units of the lattice spacing $a$,

$$\frac{d\, am_c(\mu)}{d(a\mu)^2} = \frac{\partial B_4}{\partial (a\mu)^2} \left(\frac{\partial B_4}{\partial am}\right)^{-1}. \tag{8}$$

The dependence on quark mass is rather pronounced and the second factor has been determined before [4]. All the difficulty resides in the first factor which we evaluate by calculating finite differences

$$\frac{\partial B_4}{\partial (a\mu)^2} = \lim_{(a\mu)^2 \to 0} \frac{B_4(a\mu) - B_4(0)}{(a\mu)^2}. \tag{9}$$

Because the required shift in the $\beta$'s is very small, it is adequate and safe to use the original Monte Carlo ensemble obtained for $am = am_0^c$ and $a\mu = 0$, and reweight the results by the standard Ferrenberg-Swendsen [10] method. By reweighting to imaginary $\mu = i\mu_i$[11, 12], the reweighting factors remain real positive and close to 1. Moreover, statistical fluctuations of $B_4(a\mu)$ and $B_4(0)$ cancel to a large extent under reweighting, thus decreasing the final error in the difference. A detailed discussion and successful test of this economical way of evaluating the derivative are contained in [13].

The results of this procedure for the finite difference quotient are shown in Fig. 14. Note that the statistical errors of the different data points are all correlated because all data points are generated from the same Monte Carlo ensemble. The extrapolation to zero chemical potential gives the desired derivative. We have done extrapolations by fitting to a constant value as well as to a linear function of $\mu^2$, whose slope correspondingly gives the $\mu^4$-coefficient of $B_4$ (or, more precisely, all contributions $\mathcal{O}(\mu^4)$). Both fits give consistent extrapolations. We quote the linear fit, for which the jackknife analysis yields larger, more conservative errors:

$$\frac{\partial B_4}{\partial (a\mu)^2} = 32(6) + 17000(8000)(a\mu)^2 \tag{10}$$

The sign of the leading term is clearly positive, as was found in our previous study for $N_f = 3$ equal-mass (degenerate) quark species [5]. The subleading term, although poorly determined, tends to reinforce the effect of the leading term when a real chemical potential is turned on, again consistently with [5]. However, its magnitude indicates that the approximation represented by our truncated Taylor expansion may degrade when $|a\mu| \gtrsim 0.04$, or $\mu/T \gtrsim 0.16$.

Finally, we need to supply $\partial B_4/\partial am$, plug it into Eq. (8) and convert to continuum units. A detailed prescription for this is given in [5], here we merely

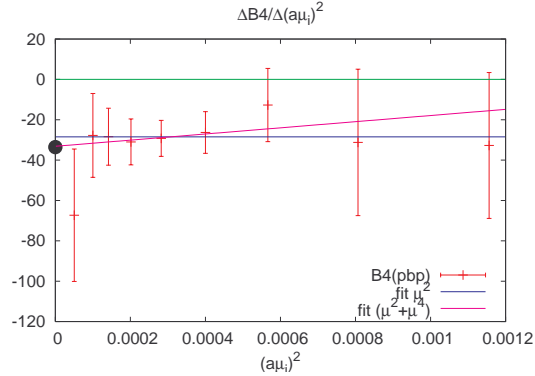

Figure 14: Change of the Binder cumulant with small imaginary chemical potential. The extrapolation to zero chemical potential gives the derivative $\partial B_4/\partial (a\mu)^2$ needed in Eq. (8).

quote our final result

$$\frac{m_{u,d}^c(\mu)}{m_{u,d}^c(0)} = 1 - 39(8)\left(\frac{\mu}{\pi T}\right)^2 + \ldots \qquad (11)$$

The increase in magnitude of the leading coefficient, from 3.3(3) to 39(8) as the strange quark mass increases from the $N_f = 3$ mass-degenerate case [5] to its physical value, is mostly due to our choice of dividing by $m_{u,d}^c(0)$ in the definition of the curvature in Eq. (11). Restoring this factor, which is $0.0265 a^{-1}$ for $N_f = 3$ and $0.005 a^{-1}$ here, gives a slowly-varying curvature.

In conclusion, we find that the critical light quark mass $m_{u,d}^c$ decreases under a small real chemical potential, just as was the case for the mass degenerate case, $N_f = 3$. This confirms that, on coarse $N_t = 4$ lattices, the chiral critical surface for small chemical potentials indeed behaves as in Fig. 1 (right). Thus, a small chemical potential weakens any phase transition. For physical quark masses the QCD transition is a crossover at $\mu = 0$ which becomes even softer as a small chemical potential is switched on. It is now most interesting and important to repeat these calculations on finer lattices, in order to see whether this behaviour of the chiral critical surface is also realized in continuum QCD.

# 7 Conclusions

This project demonstrates that the computing Grid can be used efficiently, yielding $\mathcal{O}(10^3)$ speedup to produce complex scientific results for certain types of applications. Our lattice QCD application, besides being of the embarrassingly parallel type, had the following desirable features: a large granularity (one iteration took over an hour), a small I/O requirement (10MB per hour or less), and a robust single-CPU code. These features are not typical of other lattice

QCD applications, which often simulate too many degrees of freedom to be handled by a single CPU.

For our application, the management and scheduling of $\mathcal{O}(10^3)$ independent job threads could be advantageously handled by Ganga and DIANE – user-level overlay tools above regular Grid services. With limited high-level scripting, we plugged into the Master service scheduling algorithms which exploited the knowledge of the internal structure of the QCD application. Dynamic resource selection based on the application feedback was automatically handled by AgentFactory, which allowed to reduce wasted resources to $\mathcal{O}(10\%)$. With the exception of external events such as service power outage or minor manual interventions such as upgrade of the application code or renewal of user credentials to the Grid, the system operated autonomously for several months and showed exceptional stability.

In our case, the computing Grid has made it possible to obtain scientific results faster and at a lower cost that in a High Performance Computing (HPC) center. While comparable CPU resources are commonly available there, they are usually packaged within a massively parallel machine, equipped with an expensive high-bandwidth interconnection network. While for our application the initial pre-thermalization phase was conducted using a supercomputer, for the bulk of the simulations a very large pool of loosely connected, heterogeneous PCs provides an adequate, cheaper platform.

Finally, we recall the physics question answered here. Our study shows that there is no QCD chiral critical point at temperature $T$ and small quark chemical potential $\mu_q$ satisfying $\mu_q/T < \mathcal{O}(1)$, on a coarse lattice with 4 time-slices, with a strange quark having its physical mass. This work completes the earlier study of [5], which concerned the theory with 3 quark flavors of equal, degenerate masses. It remains to be seen whether these results persist on a finer lattice.

Increasing the lattice size for our application would require an efficient strategy of handling $\mathcal{O}(100)$ distributed simulation threads each being a locally-parallel job with $\mathcal{O}(30)$ processes. The Ganga/DIANE system could readily be extended to use multicore processors or other parallel architectures available in computational grids while exploiting existing scheduling and resource selection strategies described in this paper.

# Acknowledgements:

We acknowledge the usage of EGEE resources (EU project under contracts EU031688 and EU222667). Computing resources have been contributed by a number of collaborating computer centers, most notably HLRS Stuttgart (GER), NIKHEF (NL), CYFRONET (PL), CSCS (CH) and CERN.

We would like to thank Andrew Maier and Patricia Mendez for their involvement in the early phases of the project.

# References

[1] F. Karsch, E. Laermann, and C. Schmidt. The chiral critical point in 3-flavor QCD. *Phys. Lett. B*, 520:41, 2001.

[2] P. de Forcrand and O. Philipsen. The QCD phase diagram for three degenerate flavors and small baryon density. *Nucl. Phys. B*, 673:170, 2003.

[3] S. Kim, Ph. de Forcrand, S. Kratochvila, and T. Takaishi. The 3-state potts model as a heavy quark finite density laboratory. *PoS LAT2005*, page 166, 2006.

[4] P. de Forcrand and O. Philipsen. The chiral critical line of $N_f = 2 + 1$ QCD at zero and non-zero baryon density. *JHEP*, 0701:077, 2007.

[5] P. de Forcrand and O. Philipsen. The chiral critical point of $N_f = 3$ QCD at finite density to the order $(\mu/T)^4$. *JHEP*, 0811:012, 2008.

[6] C. Germain-Renaud, C. Loomis, J.T. Mościcki, and R. Texier. Scheduling for responsive Grids. *J. Grid Computing*, 6:15–27, 2008.

[7] J.T. Mościcki, F. Brochu, J. Ebke, U. Egede, J. Elmsheuser, K. Harrison, R.W.L. Jones, H.C. Lee, D. Liko, A. Maier, A. Muraru, G.N. Patrick, K. Pajchel, W. Reece, B.H. Samset, M.W. Slater, A. Soroko, C.L. Tan, D.C. van der Ster, and M. Williams. Ganga: A tool for computational-task management and easy access to Grid resources. *Computer Physics Communications*, 180(11):2303 – 2316, 2009.

[8] P. de Forcrand and O. Philipsen. The curvature of the critical surface $(m_{ud}, m_s)^{\text{crit}}(\mu)$: a progress report. *PoS LATTICE2008*, page 208, 2008.

[9] T. Mitchel. *Machine Learning*. McGraw Hill Higher Education, 1997.

[10] A. M. Ferrenberg and R. H. Swendsen. Optimized Monte Carlo analysis. *Phys. Rev. Lett.*, 63:1195, 1989.

[11] Andre Roberge and Nathan Weiss. Gauge theories with imaginary chemical potential and the phases of QCD. *Nucl. Phys.*, B275:734, 1986.

[12] Philippe de Forcrand and Owe Philipsen. The QCD phase diagram for small densities from imaginary chemical potential. *Nucl. Phys.*, B642:290–306, 2002.

[13] P. de Forcrand, S. Kim, and O. Philipsen. A QCD chiral critical point at small chemical potential: is it there or not? *PoS LAT2007*, page 178, 2007.